\documentclass[sn-basic]{sn-jnl}
\usepackage{float}
\usepackage{graphicx}%
\usepackage{multirow}%
\usepackage{amsmath,amssymb,amsfonts}%
\usepackage{amsthm}%
\usepackage{mathrsfs}%
\usepackage[title]{appendix}%
\usepackage{soul}
\usepackage{textcomp}%
\usepackage{manyfoot}%
\usepackage{booktabs}%
\usepackage{algorithm}%
\usepackage{algorithmicx}%
\usepackage{algpseudocode}%
\usepackage{listings}%
\usepackage{multicol}
\usepackage{multirow}
\usepackage{url}
\usepackage{tabularx}
\usepackage{tabularray}
\usepackage{placeins}

\usepackage{lipsum}
\usepackage{caption}
\usepackage{cuted}
\usepackage{hyperref}
\usepackage{algorithm}
\usepackage{algpseudocode}
\usepackage{tabularx}
\usepackage{tabularray}
\usepackage{placeins}
\usepackage{gensymb}
\usepackage{multicol}
\usepackage{multirow}
\usepackage{url}

\theoremstyle{thmstyleone}%
\theoremstyle{thmstyletwo}%

\theoremstyle{thmstylethree}%

\usepackage{soul} 

\begin{document}

\title{Enhancing Diabetic Retinopathy Classification Accuracy through Dual Attention Mechanism in Deep Learning}

\author[1]{\fnm{Abdul} \sur{Hannan}}\email{khawajahannan@yahoo.com}

\author[1]{\fnm{Zahid} \sur{Mahmood}}\email{zahid0987@cuiatd.edu.pk}

\author[2]{\fnm{Rizwan} \sur{Qureshi}}\email{fnu.rizwan@ucf.edu}

\author*[3]{\fnm{Hazrat} \sur{Ali}}\email{hazrat.ali@live.com}

\affil*[1]{\orgdiv{Department of Electrical and Computer Engineering}, \orgname{COMSATS University Islamabad}, \orgaddress{\street{Street}, \city{Abbottabad}, \postcode{22060}, \country{Pakistan}}}

\affil[2]{\orgdiv{Center for Research in Computer Vision}, \orgname{The University of Central Florida}, \orgaddress{\city{Orlando}, \state{Florida}, \country{USA}}}

\affil[3]{\orgdiv{Division of Computing Science and Mathematics}, \orgname{University of Stirling}, \orgaddress{\city{Stirling}, \postcode{FK94LA}, \country{United Kingdom}}}

\abstract{Automatic classification of Diabetic Retinopathy (DR) can assist ophthalmologists in devising personalized treatment plans, making it a critical component of clinical practice. However, imbalanced data distribution in the dataset becomes a bottleneck in the generalization of deep learning models trained for DR classification. In this work, we combine global attention block (GAB) and category attention block (CAB) into the deep learning model, thus effectively overcoming the imbalanced data distribution problem in DR classification. Our proposed approach is based on an attention mechanism-based deep learning model that employs three pre-trained networks, namely, MobileNetV3-small, Efficientnet-b0, and DenseNet-169 as the backbone architecture. We evaluate the proposed method on two publicly available datasets of retinal fundoscopy images for DR. Experimental results show that on the APTOS dataset, the DenseNet-169 yielded 83.20\% mean accuracy, followed by the MobileNetV3-small and EfficientNet-b0, which yielded 82\% and 80\% accuracies, respectively. On the EYEPACS dataset, the EfficientNet-b0 yielded a mean accuracy of 80\%, while the DenseNet-169 and MobileNetV3-small yielded 75.43\% and 76.68\% accuracies, respectively. In addition, we also compute the F1-score of 82.0\%, precision of 82.1\%, sensitivity of 83.0\%, specificity of 95.5\%, and a kappa score of 88.2\% for the experiments. Moreover, in our work, the MobileNetV3-small has 1.6 million parameters on the APTOS dataset and 0.90 million parameters on the EYEPACS dataset, which is comparatively less than other methods. The proposed approach achieves competitive performance that is at par with recently reported works on DR classification.}


\maketitle

\section{Introduction}\label{sec:introduction}

Diabetes Mellitus (DM) is a long-lasting sickness in which blood sugar levels increase due to the inability of the pancreas to secrete sufficient blood insulin \cite{r1, dia}. The International Diabetes Federation (IDF) reports that over 537 million individuals between the ages of 20 and 79 worldwide have DM \cite{r2}. Over the past two decades, the prevalence of diabetes among adults has increased over threefold, and approximately 230 million people with diabetes, which is half of all cases, remain undiagnosed. An alarming estimate is that the global DM population is expected to increase to 643 million by the year 2030 and further to 783 million by 2045 \cite{r2}.

Diabetic retinopathy (DR) is a complication of diabetes that affects small blood vessels in the retina, causes blockage and bleeding in the retina, and ultimately leads to vision loss. Moreover, DR can also affect human organs like the liver, heart, kidneys, and eyes. Therefore, early detection and treatment of the DR are crucial to prevent permanent damage. The primary abnormal characteristics observed in the DR consist of (i) microaneurysms, which is an early stage of the DR in which tiny red dots appear on the retina~\cite{d2}. These dots are defined by sharp margins with a size of at most {125} $\mu$m. (ii) Hemorrhages, which are indicated by large spots on the retina surface with irregular margin sizes above 125 $\mu$mm. (iii) Hard exudates, which appear as a result of plasma leakage, become visible as yellow spots on the retina surface and span the outer retina layers with sharp margins. (iv) Soft exudates, which appear as a result of swelling of nerve fibers, have an oval-like appearance on the retina surface. Based on the nature and quantity of abnormalities observed in fundus images, the DR can be categorized into five distinct phases, which include the absence of (DR0), minor (DR1), moderate (DR2), severe (DR3), and proliferative (DR4) \cite{r2}. 
\par Deep learning methods have shown great promise in automatically classifying the distinct phases of diabetic retinopathy (DR), particularly through tailored CNN architectures~\cite{d3}. In recent years, Convolutional Neural Networks (CNNs) have made significant strides and found extensive applications in computer vision tasks such as image classification, object detection, and semantic segmentation~\cite{r3, r4, r5, r6, r7}. These networks seamlessly combine feature extraction and classification into a unified process, leading to remarkable advancements in these domains.
 Moreover, CNNs have seen extensive utilization in the domain of retinal image analysis, thus playing an important role in tasks such as blood vessel segmentation~\cite{d3}, identifying optic disc boundaries, and glaucoma screening \cite{r8}, \cite{r9, r10}. While numerous deep learning methods have been introduced for diabetic retinopathy classification on balanced datasets, however, they often struggle to address the challenge of imbalanced datasets. Improving accuracy on imbalanced datasets poses a significant challenge for algorithms that demonstrate impressive accuracy on a balanced dataset, and are over-parameterized. 

Recently, two-stage CNN architecture to detect lesions and grade DR in fundus images has shown promising results~\cite{r11}. This method addresses the DR grading process through a two-stage framework; however, the pipeline entails greater intricacy than a single-stage strategy. While CNN-based methods for grading DR have demonstrated favorable outcomes, their implementation in clinical practice remains challenging due to the inherent complexity of the task. One major difficulty arises from the striking similarities in color and texture among the five DR grades, making it prone to confusion during the grading process. This significantly hampers the performance of the model in making a distinction between the different classes. Furthermore, certain lesions within fundus images are minute, comprising only a few pixels. Hence, there is a need to explore newer methods that can achieve the task of DR classification and also effectively address the challenge of an unbalanced dataset. In this connection, we introduce an attention-based mechanism that uses pre-trained models as a backbone to perform the tasks of DR classification in retinal fundus images. Our main contributions are listed below.

\begin{itemize}
    \item We introduce a CNN-based approach while reducing the number of parameters, which automatically classifies the DR. The proposed method is trained and tested on two well-known publicly available APTOS and EYEPACS datasets. We focus on two crucial challenges. Initially, we tackle the challenge of imbalanced datasets, ensuring that appropriate accuracy is achieved. Later, we optimize the number of parameters utilized by the CNN algorithms, thereby improving their efficiency without compromising accuracy.
\item Our proposed DR classification approach incorporates the Global Attention Block (GAB) and Category Attention Block (CAB) into backbone networks, such as MobileNetV3-small, DenseNet-169, and EfficientNet-b0. The proposed approach exhibits three key characteristics for GAB and CAB. Firstly, GAB and CAB are distinct attention blocks with different functionalities. The GAB encompasses channel attention and spatial attention, while the CAB emphasizes category attention. By combining these two blocks, we enhance the performance of DR classification. Secondly, our model employs the CAB to produce class-specific attention feature maps in a category-oriented manner. Consequently, the model effectively captures more complex GAB and CAB features. This approach proves highly advantageous when dealing with imbalanced datasets in the context of DR grading tasks.

\item Our model employs the GAB (a global attention block) within a single-branch architecture specifically designed for Diabetic Retinopathy (DR) grading. The GAB captures rich global contextual features, which are subsequently refined by a Category Attention Block (CAB) that emphasizes discriminative, category-specific information. Together, these modules form a dual attention mechanism that processes spatial and channel features in parallel—unlike conventional sequential models such as CBAM and SE-ResNet. This parallel design preserves essential feature interactions and adapts dynamically to varying image contexts through a learnable unification strategy. Additionally, the proposed architecture effectively mitigates class imbalance and achieves competitive accuracy with significantly fewer parameters, as demonstrated in Section \ref{sec:discussion}, Table 3.
\end{itemize}

\begin{table}[h]
    \centering
    \caption{Key abbreviations used in the text.}
    \label{tab:table1}
    \begin{tabular}{ll}
    \toprule
        Acronym & Meaning  \\ \midrule
        CAB & Category Attention Block  \\ 
        CNN & Convolutional Neural Network  \\ 
        DNN & Deep Neural Network  \\ 
        GAB & Global Attention Block  \\ 
        FC & Fully Connected  \\ 
        DR & Diabetic Retinopathy  \\ 
        GAP & Globel Average Pooling  \\ 
        EyePACS  & Eye Picture Archive Communication System   \\ 
        APTOS  & Asia Pacific Tele-Ophthalmology Society   \\ 
        NPDR  & Non-Proliferative DR \\ 
        PDR  &  Proliferative DR  \\ 
        Acc  & Accuracy   \\ 
        MLP & Multilayer Perception  \\ 
        CBAM & Convolutional Block Attention Module  \\ 
        BNN & Bayesian Neural Network  \\ 
        PCA & Principal Component Analysis  \\ 
        \botrule
    \end{tabular}
\end{table}
The manuscript is organized as follows: Section \ref{sec:relatedwork} briefly reviews the recent literature on DR classification. Section \ref{sec:proposedmethod} presents our proposed methodology. Section \ref{sec:results} presents detailed results and discussions. Finally, conclusion is provided in Section \ref{sec:conclusion}. To facilitate the readers, Table \ref{tab:table1} shows the key acronyms used in the text. 

\section{Related Work}
\label{sec:relatedwork}

Earlier works on DR classification were mostly based on hand-crafted features~\cite{reddy2020ensemble, bhatia2016diagnosis}. These approaches typically analyzed anatomical features such as blood vessels and the optic disc to identify abnormalities like microaneurysms, hemorrhages, and exudates. However, these methods were labor-intensive and time-consuming. With the rise of Machine Learning (ML) and Deep Learning (DL), there has been a shift towards automatic feature extraction and classification. Several studies~\cite{r12, r13, r14} have leveraged deep convolutional neural networks (DCNNs) for DR detection and classification. Similarly, methods proposed in~\cite{r17, r18} focused on segmenting retinal vessels, while others~\cite{r19, iqbal2018, ahmad2022} used generative adversarial networks (GANs) to synthesize realistic retinal images for improved training.

~\cite{r20} introduced a multi-sieving CNN with an “image-to-text mapping” mechanism to locate microaneurysms in retinal images in real-time. Jain et al.~\cite{r21} assessed the effectiveness of transfer learning using VGG19, VGG16, and Inception-v3 for both binary and multi-class classification. Zeng et al.~\cite{r22} designed a Siamese CNN architecture with transfer learning to classify fundus images into two categories. The study in~\cite{r23} utilized a multilayer perceptron (MLP) alongside an enhanced Xception network, which fused multiple feature maps for better accuracy. Another approach~\cite{r24} developed four separate Inception models, each responsible for grading a quadrant of the fundus image, enabling more granular classification.

Multi-model frameworks have also been explored to improve classification performance. For instance,~\cite{r25} used a two-stage attention mechanism with ResNet50 for detecting DR and diabetic macular edema (DME), training on both Messidor and IDRID datasets. A three-stage approach was proposed in~\cite{r26}, involving preprocessing, feature extraction using ResNet50, and classification using support vector machines (SVM) and neural networks on the ISBI'2018 IDRID dataset~\cite{balagurunathan2021lung}. Similarly,~\cite{r27} presented a multi-stream deep neural network combining ResNet50, DenseNet121, PCA for dimensionality reduction, and ensemble classification using boosting techniques.

It is well established that increasing CNN depth without sufficient training data can lead to overfitting~\cite{r15, r16}. To address this,~\cite{r28} proposed a DenseNet-169 model enhanced with a Convolutional Block Attention Module (CBAM), where features were extracted, refined, and averaged using global average pooling before classification. In~\cite{r29}, a two-stage detector was used to identify specific DR lesions like hemorrhages and exudates, while Guo et al.~\cite{r30} focused on segmentation-based classification.

A comparative study in~\cite{r31} evaluated CNN, DNN, and BNN models using a dataset of 2000 retinal images for 5-stage DR classification. The DNN model achieved the highest accuracy of 86.3\%. Another work~\cite{r32} employed ResNet-34 with preprocessing techniques on the Kaggle dataset, achieving 85\% accuracy. 
~\cite{haq2024yolo} developed a hybrid YOLOv2 and ResNet-based framework achieving over 96\% accuracy in detecting and counting colorectal cancer cells, demonstrating the strength of integrated localization–classification pipelines in biomedical image analysis.

Recent advancements have focused on secure and collaborative learning strategies. Bhulakshmi and Rajput proposed the FedDEO and FedDL models~\cite{bhulakshmi2024privacy, bhulakshmi2024feddl}, leveraging federated learning to ensure privacy-aware, distributed DR classification. These approaches align with our current study's objective to develop generalized, interpretable, and privacy-preserving DR classification models.

\section{Proposed Method}
\label{sec:proposedmethod}
\begin{figure}[H]
    \centering
    \includegraphics[width=1\textwidth, height=10cm]{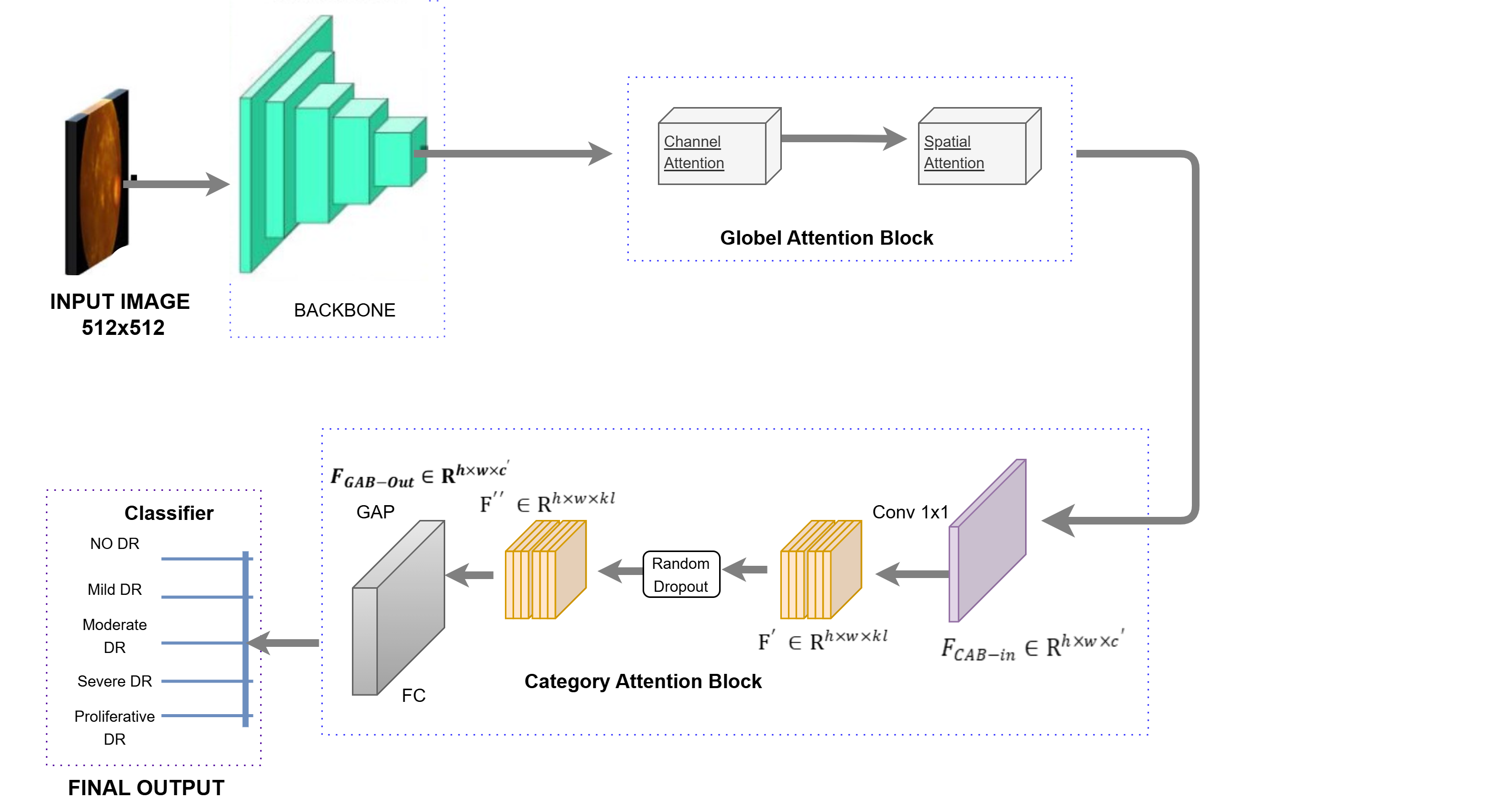}
    \caption{Overall workflow of the proposed approach for diabetic retinopathy classification.}
    \label{fig1}
\end{figure}

In this section, we present our proposed architectural design, training configurations, and the evaluation criteria employed in our study. We also explain the exploration of data augmentation, balancing techniques, and subsequent detailed analysis of our approach. Figure {\ref{fig1}} shows the overall workflow of our approach. 

\subsection{Data Preprocessing}
To enhance the accuracy of the DR classification, we have augmented the dataset using various techniques. Data preprocessing is done earlier than data manipulation to fit the data for the network that is employed in later stages. We performed the data preprocessing in the following manner.\par
\textbf{Image Rescaling:} The APTOS and EYEPACS datasets contain images of different sizes. Therefore, we rescale each image to a resolution of 512×512 pixels. \par
\begin{figure}[H]
    \centering
    \includegraphics[width=0.9\linewidth]{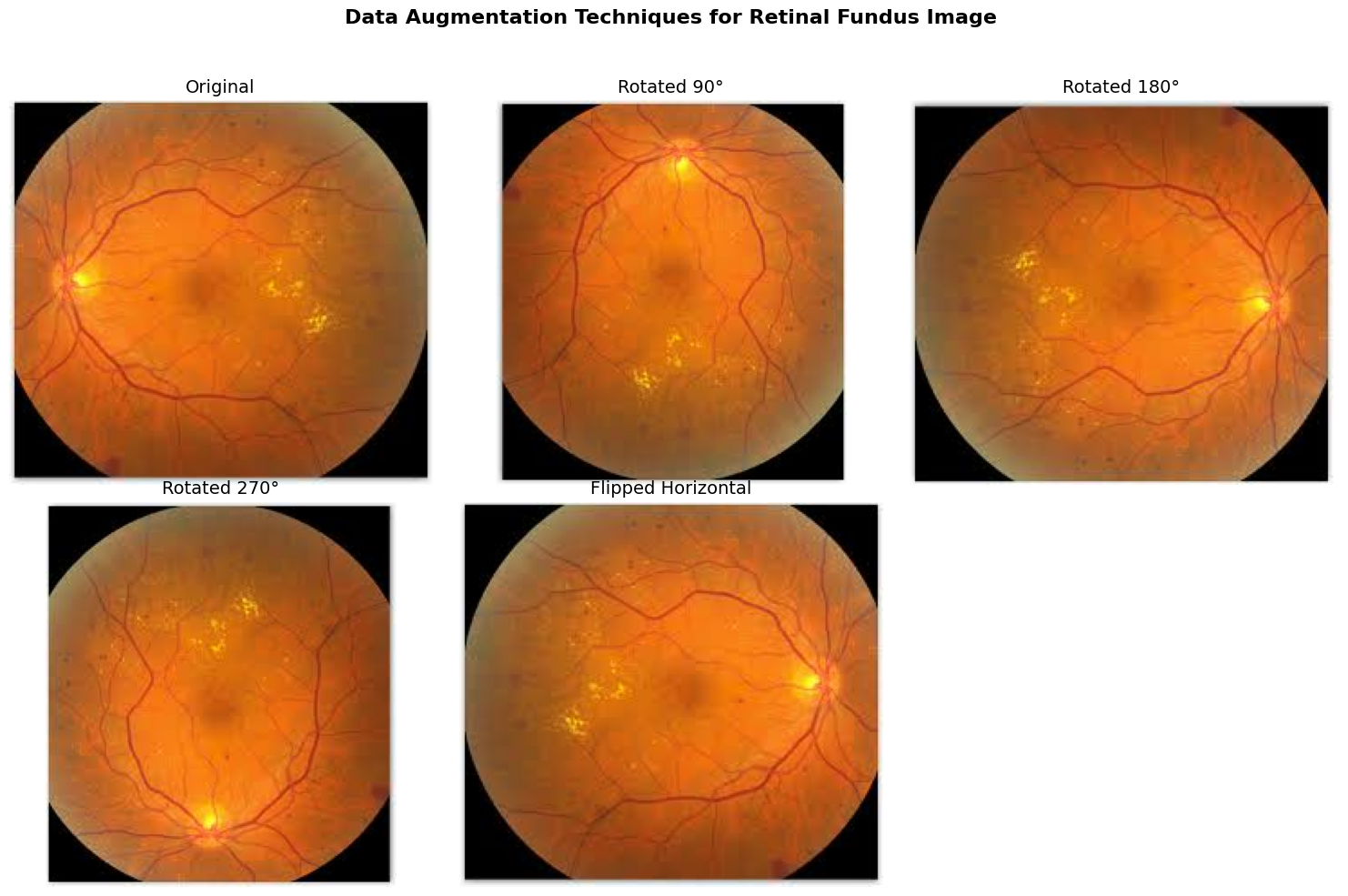}
    \caption{Augmented Image: This figure illustrates the data augmentation techniques applied to retinal fundus images, including rotations (90°, 180°, 270°) and horizontal flipping. These augmentations enhance dataset diversity and improve model generalization.}
    \label{fig:augmented_image}
\end{figure}

\textbf{Image Augmentation:} Figure 2 show the outcomes for the data augmentation techniques. To enhance the robustness and generalization capability of the model, we applied data augmentation techniques to both the APTOS and EyePACS datasets. Specifically, images were augmented using rotation (90°, 180°, and 270°) and horizontal flipping. These transformations increased the diversity of the training data and approximately doubled the dataset size by generating new, distinct variations of the original images. This augmentation strategy helps reduce overfitting and improves the model's performance in recognizing retinal abnormalities under different orientations.

\textbf{Image Labeling:} With the assistance of professional ophthalmologist, we manually label the dataset as DR0, DR1, DR2, DR3 and DR4 for DR categorized into five distinct phases. These labels refer to the absence of DR (DR0), minor DR (DR1), moderate DR (DR2), severe DR (DR3), and proliferative DR (DR4).

\subsection{Architecture}
As indicated in Figure {\ref{fig1}}, our proposed pipeline incorporates the GAB and CAB into the three backbone networks, namely MobileNetV3-small, DenseNet-169, and EfficientNet-b0. It exhibits three key characteristics for the GAB and CAB. Firstly, the GAB and CAB are distinct attention blocks with different functionalities. The GAB encompasses channel attention and spatial attention, while the CAB emphasizes category attention. Integrating these two blocks, we can enhance the performance of the DR classification algorithm. Secondly, our model employs the CAB to produce class-specific attention feature maps in a category-oriented manner. Consequently, the model can effectively capture more complex GAB and CAB features. We will analyze in the next section that this approach proves highly advantageous when dealing with imbalanced datasets in the context of DR grading tasks.

\begin{figure*}[h]
\centering
	\begin{center}
  \includegraphics[width=1 \textwidth,height=4cm]{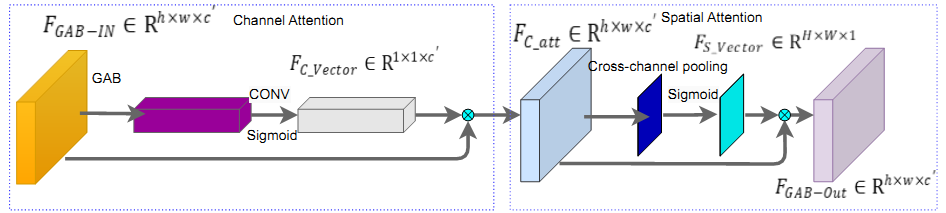}
  \caption{The GAB structure.}
  \label{fig2}
  \end{center}
\end{figure*}
Thirdly, our model incorporates a single-branch GAB, which is specifically tailored for the DR classification task. The CAB is subsequently applied utilizing the attention feature maps, which are generated by the GAB. By leveraging the enhanced global attention feature maps that are independent of specific categories, the CAB contributes significantly to the improvement of the DR grading. Below, we briefly describe the GAB and the CAB modules.\\
\textbf{The GAB:} The GAB module contains channel attention and spatial attention blocks as indicated in Figure \ref{fig2}. The GAB takes the features \(F_{reduce} \in R^{h \times w \times c^{'}}\) as input and learns global attention feature maps. Initially, we compute the channel attention feature maps using the formula specified in Eq. {\ref{eq1}}:
\begin{equation}\label{eq1}
    F_{ch\_att}=(\sigma(Conv2(GAP(F_{GAB-in}))))\otimes F_{GAB-in}
\end{equation}
Where \(F_{ch\_ att} \in R^{h \times w \times c^{'}}\), \(Conv2\) shows
two 1×1 convolutional layers, \(\sigma\) denotes the sigmoid function,
\(GAP\) denotes the pooling layer, \(F_{GAB - in}\) =
\(F_{reduce},\ \)and $\otimes$ denotes element-wise multiplication.

After this, we calculate the output of GAB. The spatial attention
feature maps \({(F}_{GAB - out}\)) are computed using Eq. {\ref{eq2}}:
\begin{equation}\label{eq2}
F_{GAB-out}=F_{ch\_att}\otimes(\sigma(C\_GAP(F_{ch\_att})))
\end{equation}
where \(C\_ GAP\) denotes the average pooling of cross channel and
\(F_{GAB-out}\ \) is the output produced by the category attention feature maps \cite{r36, zafar2024single}.\par
Figure \ref{fig2} also shows the Channel Attention and the Spatial Attention blocks. The former serves as a feature selector at the channel level, learning channel-wise attention weights to determine the significance of each feature channel and suppress less informative channels. While the latter highlights the importance of individual spatial positions by learning spatial attention weights. This aspect complements the channel attention mechanism. As indicated in Figure {\ref{fig1}}, the GAB is placed before the CAB so that we can achieve a sequential arrangement. This arrangement aims to extract comprehensive lesion information globally and preserve smaller lesion regions, thereby minimizing the information loss. The CAB focuses more on discriminative regions and enhances the features generated by the GAB. Our study indicates that reversing the order of these two blocks causes the CAB to lose fine-grained details, resulting in negative consequences for the final outcomes.\\
\textbf{The CAB:} The CAB block is constructed to learn discriminative
regions from fundus images, enhancing the fine-grained task of
classification of the DR with \(L\) numbers of classes. Considering an
incoming feature map denoted as \(F_{CAB - in}\) $\in$
\(R^{h \times w \times c}\), it is initially processed using a 1×1 convolutional layer, which yields feature maps represented as F $\in$ \(R^{h \times w \times kL}\).\par
The \(k\) represents the number of channels needed to identify distinctive regions for individual class levels. To guarantee that each of the \(k\) feature maps within a class captures unique discriminative regions, we introduce a dropout mechanism while the model is being trained. Specifically, we randomly eliminate half of the features, setting their values to zero, yielding \(F\) $\in$ \(R^{h \times w \times kL}\), which preserves half of the elements from every feature map during the dropout process. Nonetheless, during the
inference phase, the dropout operation is omitted, and all elements in the k feature maps are employed for prediction. Consequently, the scores
\(S_{i}\) \(= \ \{ S1,\ S2,...,\ SL\}\) for each class can be computed
using Eq. {\ref{eq3}}.
\begin{equation}\label{eq3}
    S_{i}=\frac{1}{k}\sum_{j=1}^{k}GMP(f'_{ij}),i\in {1,2,..,L}
\end{equation}

To obtain the feature map for each class, a category-wise cross-channel
average pooling operation is performed on \(F^{'}\). This operation
calculates the average of the feature maps across channels separately
for each class as indicated by Eq. {\ref{eq4}}.
\begin{equation}\label{eq4}
    F'_{i\_avg}=\frac{1}{k}\sum_{j=1}^{k}(f'_{ij}),i\in {1,2,..,L}
\end{equation}
Where \({f'}_{i,j}\) denotes the representation of the ith class
pertaining to the jth feature map extracted from \(F^{'}.\) Moreover the
term \({F'}_{i\_ avg}\) $\in$ \(R^{h \times w \times 1}\) are semantic feature maps specific to the ith class category attention
\(\ {ATT}_{CAB}\) $\in$ \(R^{h \times w \times 1}\) that can be obtained by:

\begin{equation}\label{eq5}
    ATT_{CAB}=\frac{1}{L}\sum_{i=1}^{L}S_{i}
\end{equation}
Where the Category Attention Block (\({ATT}_{CAB})\) highlights the
informative regions for DR grading. Finally, the feature maps input of
category attention block \({F}_{CAB - in}\) can be converted to the
feature maps output of category attention block \({F}_{CAB - out}\) by
the \({ATT}_{CAB}\) as indicated by Eq. {\ref{eq6}}.
\begin{equation}\label{eq6}
  F_{CAB-out}=F_{CAB-in}\otimes ATT_{CAB}  
\end{equation}
Where \({F}_{CAB - out}\) represents the resulting feature maps from
CAB, which enhances the discerning areas within \({F}_{CAB - in}\) to
determine the severity of the DR.
The proposed CAB addresses the issue of imbalanced dataset distribution in DR grading datasets. In conventional CNNs, all feature maps are combined without considering the distinct categories. This may result in mixed information among different categories with less focus on categories with fewer samples of feature maps. To overcome this, the CAB allocates a specific number of feature channels to each DR category, confirming that each DR grade receives an equal number of feature channels. This approach helps prevent channel prejudice and increases the differentiation between different DR categories. As a result, CAB effectively mitigates the problem of imbalanced data distribution commonly encountered in DR grading datasets. The CAB efficiently produces attention features using a limited number of parameters that lead to reduced computational cost. It consolidates the distinctive regions related to each category into a single feature map, which shares the same dimensions as the original feature maps.
As a result, integrating CAB with the GAB module becomes feasible, thus allowing for a unified combination with significant performance. Algorithm 1 shows the pseudo-code of our approach. From lines (2) to (8), data preprocessing is done in which each image is manually labeled and rescaled into a resolution of 512×512 pixels. Later augmentation and data manipulations are also performed, as indicated in lines (7) and (8), respectively.\par
Lines (9) to (18) in Algorithm 1 indicate the selection and application of backbones and several training strategies. Line (10) indicates the loading of three networks, MobileNetV3, Efficientnet-b0, and DenseNet-169, as a backbone model. Meanwhile, pre-trained layers are frozen, trained, and then unfrozen before adding the GAB and the CAB module. Later Eq. (1) and (2) are used to select and highlight the features through the GAB module. To address the class imbalances, Eq. \ref{eq6} is used through the CAB module. Lines (19) through (26) describe the transfer learning and other network parameters. In this step, the number of epochs and the batch size are set to 40 and 16, respectively. Meanwhile, for better CNN training, two different learning rates are set, as depicted in lines (23) and (24) of Algorithm 1, along with the selection of five channels.
\begin{algorithm}
\caption{Pseudocode of proposed DR classification method}\label{alg}
\begin{algorithmic}[1]
\item
  \textbf{Input:} Obtain colored DR images from Kaggle
\item
  \textbf{do:}
\item
  \textbf{Process} collected data obtained in line (1).
\item
  Perform pre-processing operations:
\item
  Manually label and rescale each image to a resolution of 512×512
  pixels.
\item
  Perform image augmentation and rotate images at 90\textsuperscript{0},
  180\textsuperscript{0}, and 270\textsuperscript{0}.
\item
  Perform data manipulation. $\blacktriangleright$ Isolate into validation, train, and test data set
\item
  \textbf{end}
\item
  \textbf{begin}
\item
  Load MobileNetV3-small OR Efficientnet-b0 OR DenseNet-169, as a backbone model.
\item
  Select backbone model
\item
  Freeze the pre-trained layers and train upper layers
\item
  Unfreeze the pre-trained layers.
\item
  Add Conv 1x1 Layer
\item
  Add the GAB and the CAB modules
\item
  Use Eq. \ref{eq1} and Eq. \ref{eq2} to highlight and select the features by the
  GAB.
\item
  Use Eq. (6) addressing the issue of imbalanced dataset by CAB.
\item
  \textbf{end}
\item
  \textbf{begin}
\item
  Transfer learning of the CNN
\item
  Apply fine tuning and set the parameters as:
\item
  Epochs = 40 and Batch size = 16,
\item
  Set learning rate 1 = \(5\  \times 10^{- 3}\).
\item
  Set learning rate 2 = 8\(\  \times 10^{- 5}\)
\item
  No of channels \(K = 5\)
\item
  \textbf{end}
\item
  Apply FC and the GAP classifier:
\item
  \textbf{Output:} Final classification result:
\item
  DR \{Absent, Mild, Moderate, Severe, or Proliferative\}
\end{algorithmic}
\end{algorithm}

In line (27), the Fully Connected (FC) and the GAP classifiers are applied to classify the input image into one of the five classes that we refer to as DR0, DR1, DR2, DR3, and D4, which indicate No DR, Mild DR, Moderate DR, Severe DR, and Proliferative DR, respectively. The complete steps are shown in Algorithm 1, which depict the full flow of our approach. In the next section, we analyze the performance of our approach, along with discussions and comparisons.

\section{Results}
\label{sec:results}
In this section, we briefly describe the system descriptions, training settings, datasets used, and evaluation parameters used in this study. We also analyze the performance of our approach on two publicly available datasets and present the discussions and comparisons.
\subsection{System Specifications and Training Settings}
We ran the experiment on a Google Colab GPU T4, 25 GB of RAM. Our method employs a CNN-based model and an attention module. As described above, we also used other architectures, for instance, MobileNetV3-small, DenseNet-169, and EfficientNet-b0, as a backbone model to see the outcomes yielded by our method. These models have fewer parameters. We also fine-tuned the network to deliver efficient features. Our dataset was divided into training, validation, and testing sets, with portions of 50\%, 30\%, and 20\%, respectively. The network's input resolution is 512$~\times~$512 pixels. We started with an initial learning rate of \(5 \times 10^{- 3}\). For the non-satisfactory model's improvement on the validation set for three consecutive epochs, we reduced the learning rate by a factor of 0.8. Training continued for 40 epochs using the Adam optimizer and the cross-entropy loss function. The batch size was set to 32 images per batch.

\subsection{Datasets}
We used APTOS and EYEPACS benchmark dataset in our experiments. Below, we briefly describe the datasets.

\textbf{APTOS Dataset:} It was launched in 2019 through the Kaggle competition \cite{r33}. This dataset included a fundus images dataset of numerous DR severity levels. Overall, the APTOS dataset consists of 3,669 images. The primary objective of using the fundus imaging dataset was to assess the severity of the disease by generating a probability score that an image belonged to one of the five categories. A usual method to grade the DR includes ranking its difficulty into five different classes, which are (i) No DR, (ii) Minor (mild) DR, (iii) Moderate DR, (iv) Severe DR, and (v) Proliferative DR.

\textbf{The EYEPACS Dataset:} It consists of 35,108 color fundus images for DR grading \cite{r34}. The images were captured under various conditions through several devices, which were provided by EYEPACS. These were placed at multiple primary care sites throughout California and other locations in the USA. We observe a class imbalance, with 73.3\% of images belonging to the No DR class, 6.9\% to the Minor DR class, 15.2\% to the moderate DR class, 2.6\% to the severe DR class, and 2\% to the Proliferative DR class. The class imbalance issue, as described above, indicates that these datasets are crucial and could be handy to investigate the performance of deep learning algorithms.

\subsection{Evaluation Metrics}
In our work, the performance was assessed using four commonly used metrics as described briefly below from Eq. \ref{eq7} through Eq. \ref{eq10}.

\begin{equation}\label{eq7}
    Acc=\frac{(TP+TN)}{(TP+FP+TN+FN)}
\end{equation}
\begin{equation}\label{eq8}
    Se=\frac{(TP)}{(TP+FN)}
\end{equation}
\begin{equation}\label{eq9}
    Sp=\frac{(TN)}{(TN+FP)}
\end{equation}
\begin{equation}\label{eq10}
    F1=\frac{TP}{TP+\frac{1}{2}(FN+FP)}
\end{equation}
Where \textit{Acc, Se, Sp} indicate the Accuracy, Sensitivity, and the
Specificity, respectively. The TP/TN indicates True Positives/Negatives
and FP/FN indicates False Positives/Negatives, respectively.

The quadratic weighted kappa (QWK) score for a 5-level classification of DR measures the agreement between raters. It is
calculated using the formula:
\begin{equation}\label{eq11}
    Kappa-score=K=1-\frac{\sum_{i,j}^{}W_{i,j}O_{i,j}}{\sum_{i,j}^{}W_{i,j}E_{i,j}}
\end{equation}
Where, \(W_{i,j}\)\hspace{0pt} is the pre-defined weight for agreement
between raters i and j, \(O_{i,j}\) is the observed
agreement, and $E_{i,j}$ is the expected agreement by chance. The resulting K value ranges from [-1,+1] , with $+1$
indicating excellent agreement, 0 indicating chance agreement, and
negative values indicate better than random chance agreement.

\subsection{Classification Analysis }
\begin{figure*}[h]
\centering
	\begin{center}
  \includegraphics[width=0.8 \textwidth,height=4cm]{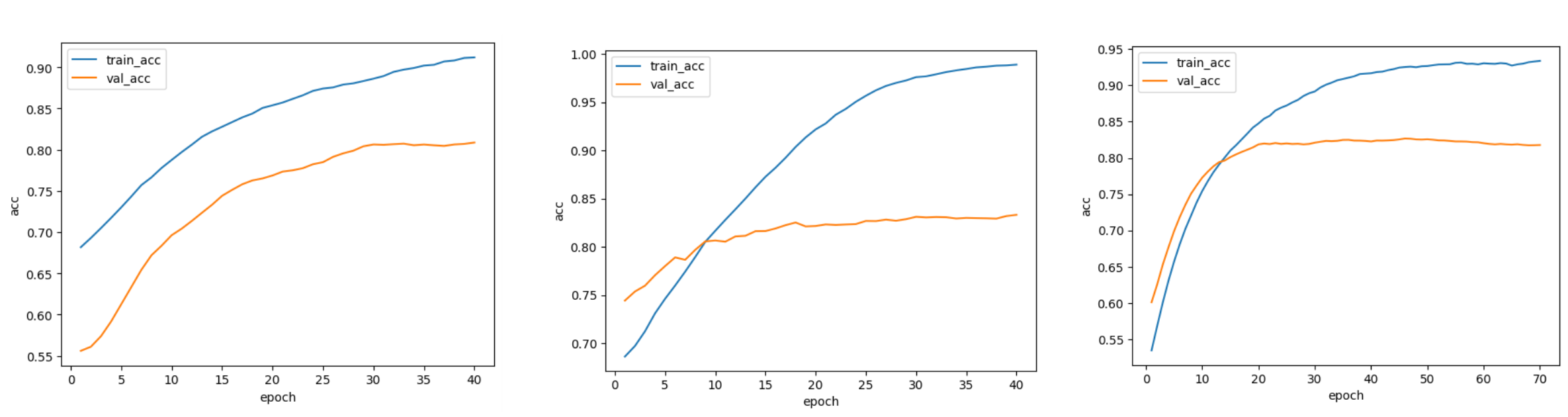}\\
 {\hspace{0cm}\small{(a). MobileNetV3-small }  }            {  \hspace{1.5cm} \small{(b). DenseNet-169}  }              {\hspace{2cm}(c). EfficientNet-b0}
  \caption{Training and validation curves for APTOS dataset, accuracy versus number of epochs}
  \label{fig3}
  \end{center}
\end{figure*}

We employed supervised learning to classify the five different classes
of the DR. Figure~\ref{fig3} shows the training curves (accuracy versus epochs) on
the APTOS dataset for MobileNetV3-small, DenseNet-169, and
EfficientNet-b0 architectures, respectively. These backbones were selected for their proven effectiveness in medical image analysis. Their lightweight architectures and feature extraction capabilities make them well-suited for retinal disease classification tasks such as diabetic retinopathy detection. For MobileNetV3-small, as
shown in Figure \ref{fig3}(a), both the training and validation accuracy
increases with the number of epochs. However, a slight decline
in validation accuracy is observed at the 38\textsuperscript{th} epoch.
Overall, the training accuracy reaches up to 93\% and validation accuracy reaches 82\%, following a monotonic rising pattern.
Similarly, Figure \ref{fig3}(b) depicts that for the DenseNet-169 network, both the
training and validation accuracies generally increase up to
10\textsuperscript{th} epoch. Here, the training accuracy increases up
to 98.5\% for 39\textsuperscript{th} epoch, while the validation
accuracy stays near constant at 84\% up to the 40\textsuperscript{th}
epoch. For the EfficientNet-b0 architecture, as shown in Figure 3(c), it is
evident that both the training and validation accuracies rise up to the
17\textsuperscript{th} epoch; however, after that, the validation
accuracy curve stays flat at 82\%, while the training accuracy
approaches nearly 93.5\%. In this case, the epochs were set to 70 to
observe the detailed trends in training and validation accuracies. The
analysis of training versus accuracy curves indicates that these models
converge up to 40 epochs.
\begin{figure*}[h]

\centering
	\begin{center}
 
  \includegraphics[width=0.8 \textwidth,height=4cm]{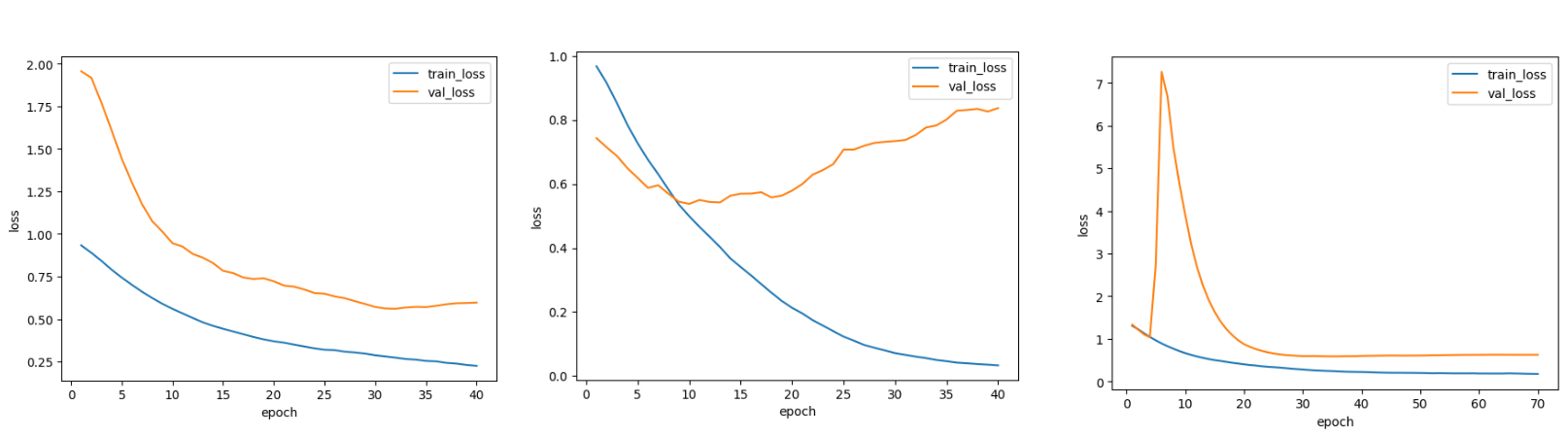}\\
 {\hspace{0cm}\small{(a). MobileNetV3-small }  }            {  \hspace{1.5cm} \small{(b). DenseNet-169}  }              {\hspace{2cm}(c). EfficientNet-b0}
  \caption{Loss curves for APTOS dataset showing training and validation losses versus number of epochs.}
  \label{fig4}
  \end{center}
\end{figure*}

\par
Figure \ref{fig4} shows the loss curves for the APTOS dataset
for MobileNetV3-small, DenseNet-169, and EfficientNet-b0 architectures,
respectively. As shown in Figure 4(a), for MobileNetV3-small, both
the training and validation losses decrease when the epochs are set to
40. Overall, it is observed that till the 40\textsuperscript{th} epoch, the
training and validation losses are found to be 0.20\% and 0.60\%,
respectively. Moreover, Figure \ref{fig4}(b) depicts that both the
training and validation losses decrease up to 10\textsuperscript{th}
epoch. For the 40 epochs that are set for this arrangement, we observe
that the training loss converges to nearly zero. For this scenario, the
validation loss was found to be 0.80. In Figure \ref{fig4}(c), the training and
validation loss for EfficientNet-b0 indicate that both the training and
validation losses are near zero. The analysis of loss versus epoch
curves presented above on the APTOS dataset suggests that all models can
converge within 40 epochs. Based on this analysis, the optimal model is
selected by identifying the minimum validation set loss. The
EfficientNet-b0 was trained for 70 epochs, resulting in an 82\%
validation accuracy. Additionally, a peak validation accuracy of 84\%
was attained using DenseNet-169 on the APTOS dataset.
\begin{figure}[H]
    \centering
    \includegraphics[width=0.9\linewidth]{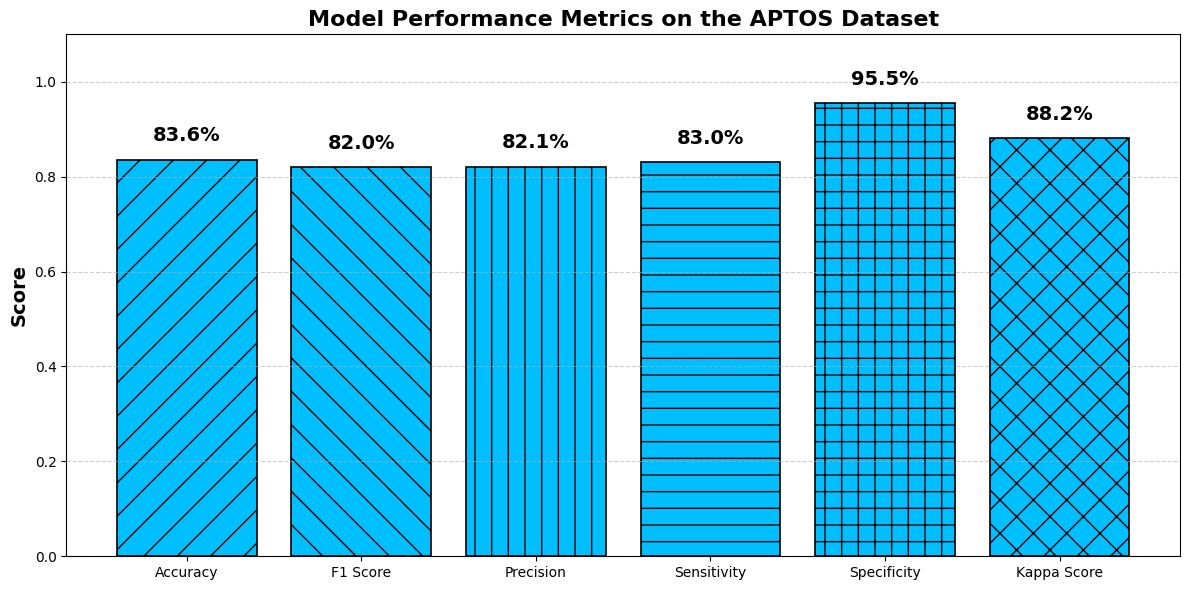}
    \caption{Model performance evaluated on the APTOS dataset. The figure illustrates key metrics including accuracy (83.6\%), F1 score, precision, sensitivity, specificity, and kappa score, using distinct hatch patterns for visual clarity. These parameters reflect the effectiveness and robustness of the proposed classification model.}
    \label{fig:parameters}
\end{figure}

Figure~\ref{fig:parameters} illustrates the key performance metrics of the proposed model evaluated on the APTOS dataset. The model achieved an overall accuracy of 83.6\%, along with an F1 score of 82.0\%, precision of 82.1\%, sensitivity of 83.0\%, specificity of 95.5\%, and a kappa score of 88.2\%. These results highlight the model’s strong and balanced ability to detect diabetic retinopathy across different severity levels. The use of distinct hatch patterns enhances visual clarity and facilitates comparison between the evaluation metrics.

\par 
The analysis presented above on the APTOS dataset gives good insight
into the performance of three different architectures.

\begin{figure}[H]
\centering
\includegraphics[width=\textwidth]{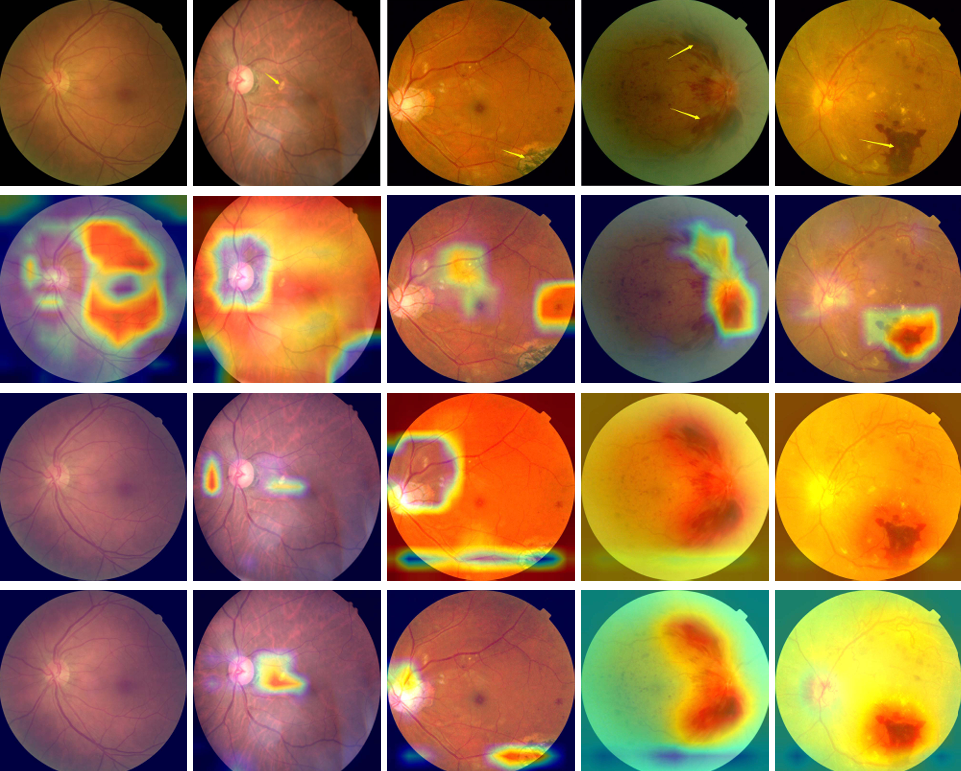}
\vspace{2mm}
\begin{tabular}{@{}>{\centering\arraybackslash}p{2.3cm}
                >{\centering\arraybackslash}p{2.3cm}
                >{\centering\arraybackslash}p{2.3cm}
                >{\centering\arraybackslash}p{2.3cm}
                >{\centering\arraybackslash}p{2.3cm}@{}}
\small\textbf{No DR} &
\small\textbf{Mild DR} &
\small\textbf{Moderate DR} &
\small\textbf{Severe DR} &
\small\textbf{Proliferative DR} \\
\end{tabular}

\caption{Grad-CAM visualization results of GAB and CAB on Aptos dataset. We show five DR grade levels (0–4 from left to right, i.e. No DR, Mild DR, Moderate DR, Severe DR and Proliferative DR, respectively). The top row provides original images where yellow arrows indicate the lesion regions. The second row provides the heatmaps without attention, the third row provides the heatmaps of GAB, and the bottom row shows the heatmaps refined by CAB.}
\label{fig:dr_gradcam}
\end{figure}

\noindent
To evaluate the explainability of the proposed model and understand the influence of CAB, we employed Grad-CAM visualizations. As illustrated in the Figure \ref{fig:dr_gradcam}, the top row presents five retinal images from the APTOS dataset, representing diabetic retinopathy levels from 0 to 4 (No DR to Proliferative DR). The second row displays attention maps generated without applying attention mechanisms, which tend to highlight irrelevant areas or miss key lesion sites—especially in DR 0, DR 1, DR 3, and DR 4. The third row shows the outputs of the GAB, which improves feature localization compared to the non-attentive model, but occasionally focuses on non-discriminative regions due to its global perspective. Finally, the fourth row includes results refined by the CAB, demonstrating clearer localization of disease-specific features and more precise coverage of lesion regions. This improvement enhances the transparency of classification decisions and supports potential clinical adoption. Notably, even subtle lesions—such as those in DR 1 are accurately highlighted by the model, indicating its robustness in detecting early-stage diabetic changes. To enhance clinical trust and support decision-making by ophthalmologists, our model incorporates attention-based interpretability techniques such as Grad-CAM, which visually highlight regions in the retinal fundus image contributing most to the model's prediction. By presenting these attention heatmaps, clinicians can cross-reference the model's focus areas with known pathological features such as microaneurysms, hemorrhages, and exudates. This visual justification not only aligns the model's decisions with clinical reasoning but also increases transparency in the diagnostic process. Furthermore, by showing the evolution of attention from generic to refined lesion-focused regions through the GAB and CAB modules, the model demonstrates a human-like reasoning pattern. Such interpretable outputs are crucial in high-stakes environments like diabetic retinopathy screening, where false negatives or overlooked regions can severely affect patient outcomes.

\subsection{Ablation Studies on APTOS Dataset}

To evaluate the effectiveness of the proposed dual attention mechanism, we conducted an ablation study using DenseNet-169 as the baseline model on the APTOS dataset. Table~\ref{tab:ablation_study} presents the comparison of different configurations in terms of accuracy, F1 score, and number of parameters. The baseline model without any attention modules achieved an accuracy of 79.0\% and an F1 score of 80.0\%. Incorporating the Global Attention Block (GAB) alone improved the performance to 82.2\% accuracy and 81.5\% F1 score, highlighting its effectiveness in capturing global contextual features. The use of only the Channel Attention Block (CAB) yielded a similar improvement, reaching 82.1\% accuracy and 81.2\% F1 score, demonstrating its capability in enhancing relevant channel-wise information. The best results were achieved when both GAB and CAB were integrated, resulting in an accuracy of 83.6\% and F1 score of 82.0\%. Importantly, the total number of parameters increased only slightly from 16.8M to 17.0M, confirming the lightweight nature of the model. This study confirms that each attention module contributes positively to performance, and their combination provides complementary advantages for diabetic retinopathy classification.
The individual contributions of the GAB and CAB modules are quantified in Table~\ref{tab:ablation_study}. When added separately to the baseline model, GAB improved the accuracy by 3.2\% and the F1 score by 1.5\%, while CAB alone improved the accuracy by 3.1\% and the F1 score by 1.2\%. The combination of both modules led to the best performance, with a total gain of 4.6\% in accuracy and 2.0\% in F1 score. These results clearly demonstrate that each attention block contributes uniquely to performance improvements, and their synergy provides complementary enhancements in diabetic retinopathy classification.
The dual attention mechanism—comprising Global Attention Block (GAB) and Channel Attention Block (CAB)—was chosen intentionally to capture both spatial and channel-wise contextual information, which are crucial for accurate diabetic retinopathy classification. GAB focuses on the global spatial structure of the lesion areas, while CAB highlights the most relevant feature channels. Although this setup adds modest complexity, the ablation study in Table~\ref{tab:ablation_study} clearly demonstrates that each module contributes uniquely and significantly to performance improvement. Compared to models with only one attention mechanism, our dual attention approach achieves better accuracy and F1 scores with minimal increase in parameters. This performance gain justifies the slightly higher architectural complexity, especially given the lightweight nature of our model, which remains suitable for real-time and low-resource applications.

\begin{table}[ht]
\centering
\caption{Ablation study of the model adopting DenseNet-169 as the baseline on APTOS dataset. Accuracy, F1 Score, and number of parameters (\# Parameters) are reported.}
\label{tab:ablation_study}
\begin{tabular}{|l|c|c|c|}
\hline
\textbf{Method} & \textbf{Accuracy (\%)} & \textbf{F1 Score (\%)} & \textbf{\# Parameters (M)} \\
\hline
Baseline (DenseNet-169) & 79.0 & 80.0 & 16.8 \\
+ GAB only              & 82.2 & 81.5 & 16.8 \\
+ CAB only              & 82.1 & 81.2 & 16.8 \\
+ GAB + CAB (Ours)      & \textbf{83.6} & \textbf{82.0} & 17.0 \\
\hline
\end{tabular}
\end{table}

\subsection{Discussion}
\label{sec:discussion}
Advancements in fundoscopy devices and deep learning
algorithms have sparked significant interest in automatic DR screening. While deep learning techniques have
demonstrated promising results in classifying the DR, there still
remains a notable disparity when it comes to their practical application in clinical settings. The improved classification accuracy of the proposed model carries significant clinical relevance. Enhanced precision—particularly the reduction of false negatives—enables earlier detection of diabetic retinopathy (DR), which is critical in preventing irreversible vision loss and ensuring timely treatment. The model’s reliability supports its use as a decision-support tool for ophthalmologists, acting as a second reader and reducing diagnostic workload in large-scale screening programs. Notably, the model is lightweight, with a reduced number of parameters compared to conventional architectures, making it highly efficient and practical for real-world deployment. Its low computational cost allows seamless integration into real-time DR screening pipelines, including on edge devices and mobile health platforms. This makes it especially suitable for use in low-resource or rural healthcare settings, where access to specialized diagnostic tools is limited. In summary, the model not only advances diagnostic accuracy but also enhances scalability, accessibility, and practical utility in diverse clinical environments. Due to its compact architecture and reduced parameter count, the model achieves faster inference speeds with minimal hardware requirements. This makes it ideal for deployment in mobile screening units, edge devices, or real-time clinical workflows without compromising diagnostic accuracy.

\begin{itemize}
\item
  This study presents the CABNet, which is a novel method specifically
  developed to tackle the issue of imbalanced data distribution during
  DR detection. To further enhance the model's interpretability, we
  additionally generate location maps that indicate potentially abnormal
  lesions in fundus images. This feature aids ophthalmologists in making
  more accurate judgments based on the model's outputs. The experiments
  demonstrate the flexibility of our approach and further show its
  effectiveness across various backbones and its ability to achieve
  superior performance on two publicly available datasets. The success
  of our approach in the DR grading can be attributed to two key
  elements, as discussed in the preceding experimental section: the
  Category Attention Block and CABNet, which is a combination of the CAB
  and the GAB.
\item
The dual attention-based model proposed in this study holds promising clinical potential for real-world diabetic retinopathy (DR) screening. Early diagnosis is vital in preventing permanent vision loss, especially in remote or underserved areas where access to ophthalmologists is limited. By effectively identifying underrepresented stages like mild and moderate DR, the model can assist in timely intercession that may otherwise be missed by standard methods. Its lightweight architecture and reduced parameter count make it suitable for mobile and teleophthalmology applications, supporting scalable and cost-effective screening initiatives. Moreover, the attention mechanisms enhance the model's transparency by accenting key retinal abnormalities such as hemorrhages and microaneurysms, making the outputs easier for clinicians to validate. These features enable the model to function as a practical assistive tool in clinical workflows, reducing diagnostic burden while improving detection coverage and patient outcomes.

\item 
The RL-MODE framework \cite{singh2024nextgen} integrates reinforcement learning with multi-objective decision-making to optimize Quality of Service metrics in resource-constrained IoT networks. This hybrid approach dynamically balances critical parameters including power efficiency, data transmission delays, and network throughput. Such adaptive optimization techniques show promising potential for managing multifaceted health data streams in gestational diabetes care, where they could enable more precise monitoring and tailored treatment strategies for pregnancy-related metabolic complications.
Recent research by \cite{singh2024ai} demonstrates how artificial intelligence-enhanced optimization techniques can strengthen security systems in consumer devices. The proposed methodology employs bio-inspired search algorithms to dynamically improve threat identification while maintaining system efficiency. These advanced computational strategies show significant potential for adaptation to healthcare diagnostics, particularly in developing more sensitive screening tools for diabetes-related reproductive health disorders through pattern recognition in complex clinical data.
The integration of AI-driven metaheuristic algorithms enhances classification accuracy, facilitating earlier and more precise detection of diabetic complications in gynecological health. This advancement contributes to timely interventions and reduces the likelihood of missed diagnoses, thereby improving patient outcomes \cite{singh2024metaheuristic}.
The AI-powered metaheuristic framework presented in this study is designed for efficient integration into real-time screening systems. Its lightweight and adaptable nature makes it suitable for deployment in low-resource healthcare settings, enhancing the timely detection of diabetic complications in gynecological health \cite{y12024}.
The survey systematically classifies facial recognition techniques into three primary categories: static image analysis, dynamic video processing, and three-dimensional modeling approaches. \cite{mahmood2017review}. Through comparative evaluation using benchmark datasets, the analysis highlights the distinct advantages and constraints of each methodology while providing practical recommendations for algorithm selection tailored to different implementation scenarios. \cite{mahmood2017review}.

\item
Despite its commendable performance, our proposed method still has areas for enhancement. To illustrate practical applicability, imagine a healthcare setting where a technician captures a fundus image using a portable retinal camera. Our proposed model processes the image and identifies moderate DR, prompting timely referral to a specialist. This scenario emphasizes the model’s potential for supporting DR screening in under-resourced environments. Firstly, the entire network is trained solely with image-level supervision, which makes it highly challenging to pinpoint smaller lesion regions accurately. Secondly, in terms of clinical application, our model can furnish a grading score and an approximation of the lesion regions. However, it does not address localization and detection tasks. Furthermore, our work does not specify the type of DR lesion, such as soft exudate, hard exudate, microaneurysm, or hemorrhage. This information is crucial for the DR screening and should be a focal point for future research endeavors.

\end{itemize}

\begin{table*}[h]
\caption{Comparison with recent works on DR classification.}
\label{tab:table2}
\renewcommand{\arraystretch}{1.5}
\centering
\resizebox{\textwidth}{!}{%
\begin{tabular}{ll|llll|llll}
\toprule
\multirow{2}*{Ref~} & \multirow{2}*{Backbone~} & \multicolumn{4}{c|}{APTOS} & \multicolumn{4}{c}{EyePACS} \\
 & & Acc \% & F1-score & Kappa-score & \#Para~ & Acc \% & F1-score & Kappa-score & \#Para~ \\
\midrule

\cite{r28}~   & DenseNet-169~     & 82~     & 0.68~    & 0.88~ & 8.50~   & –~       & –~      & –~    & –~       \\

\cite{r36}~   & DenseNet-121~     & –~      & –~       & –~    & –~      & 86\%~    & –~      & 0.86~ & 8.12 M~  \\

\cite{r37}~   & DenseNet-121~     & 72~     & –~       & –~    & 14 M~   & 85\%~    & –~      & –~    & 14 M~    \\

\cite{r38}~   & –~                & 83.50~  & –~       & –~    & –~      & –~       & –~      & –~    & –~       \\

\cite{r39}~   & –~                & 80~     & –~       & –~    & 2.20 M~ & –~       & –~      & –~    & –~       \\

\cite{r40}~   & –~                & –~      & 0.53~    & –~    & 45 M~   & –~       & –~      & –~    & –~       \\
\multirow{3}*{Proposed Method~} & MobileNetV3-small & 82~     & 0.82~    & 0.88~ & 1.6 M~  & 76.68~   & 0.72~  & 0.52~ & 0.90 M~ \\

 & DenseNet-169 & \textbf{83.60}~  & \textbf{0.82}~ & \textbf{0.88}~ & \textbf{17 M}~ & \textbf{75.43}~ & \textbf{0.73}~ & \textbf{0.59}~ & \textbf{18 M}~ \\
 & EfficientNet-b0 & 80~     & 0.79~    & 0.87~ & 7.1 M~  & 80~       & 0.73~  & –~     & 7.3 M~ \\
\bottomrule
\end{tabular}
}
\end{table*}

\subsection{Comparison}
Table \ref{tab:table2} compares our method with a few recent methods.
Below, we list important observations while comparing our methods.

\begin{itemize}
\item
  On the APTOS dataset, our proposed method, DenseNet-169 achieved the
  highest accuracy of 83.60\% and thereby outperforms all the six
  compared methods. Moreover, our proposed approach with MobileNetV3-small yields at
  par accuracy of 82\% with the work developed in \cite{r28}. The same is the
  case with \cite{r39}, where our proposed EfficientNet-b0 has a similar
  accuracy. Overall, on the APTOS dataset, the work in \cite{r37} yields
  the least accuracy than the compared methods. The mean accuracy of our proposed method by combining three architectures is
  81.86\%. Furthermore, our approach yields the highest F1-score and
  Kappa-score on the APTOS dataset. Additionally, our proposed model,
  MobileNetV3-small requires 1.6 million parameters for the APTOS
  dataset and 0.9 million for the EYEPACS dataset. On this dataset, the
  work in \cite{r40} requires the highest number of parameters,
  which is 40 million.
\item
  On the EYEPACS dataset, our proposed methods, MobileNetV3-small,
  DenseNet-169, and EfficientNet-b0 yield 76.68\%, 75.43\%, and 80\%
  accuracies, respectively. Moreover, our proposed MobileNetV3-small,
  DenseNet-169, and EfficientNet-b0 yield 76.68\%, 75.43\%, and 80\%
  accuracy, respectively. Although here, our method ranks
  3\textsuperscript{rd} in terms of accuracy, but it achieved an F1-score
  of 0.72. Further, our proposed model, MobileNetV3-small and
  EfficientNet-b0 requires 0.9 million and 7.3 million parameters,
  respectively on the EYEPACS dataset. These parameters are considerably
  less than the methods reported in \cite{r36} and \cite{r37}.
\item
  As indicated in Table 3, most of the methods either require high
  computations or have limited validation on datasets. Our work is the
  newest addition to the domain, and it uses the fine-tuned version of
  the GAB. We anticipate relatively higher DR classification accuracy
  with much less computational complexity. Similarly, a few of the
  above-described methods do not handle the class imbalance problem.
\end{itemize}

Our proposed method handles the class imbalance issue with the aim of
achieving accurate and reliable DR classification. In this research,
multiple architectures are implemented and analyzed. Therefore, we are optimistic that our developed algorithm will be robust against the variations that are
provided in the standard datasets, such as APTOS and EYEPACS. Table 3 shows that on the APTOS dataset, the proposed algorithm
outperforms MobliNetV3-small (backbone) in terms of parameters
and accuracy. The bottom line of this study is that on the
APTOS dataset, the MobileNetV3-small architecture in our proposed
algorithm requires 1.6 million parameters and outperforms \cite{r28}, \cite{r37}, \cite{r39}, and \cite{r40}. On the other hand, on the APTOS dataset, our proposed algorithm outperforms \cite{r28}, \cite{r36}, \cite{r37}, \cite{r38}, \cite{r39}, and \cite{r40} by yielding 83.6\% accuracy using the DenseNet-169 backbone.

\FloatBarrier
\begin{figure*}[]

\centering
	\begin{center}
 
  \includegraphics[width=0.8 \textwidth,height=14cm]{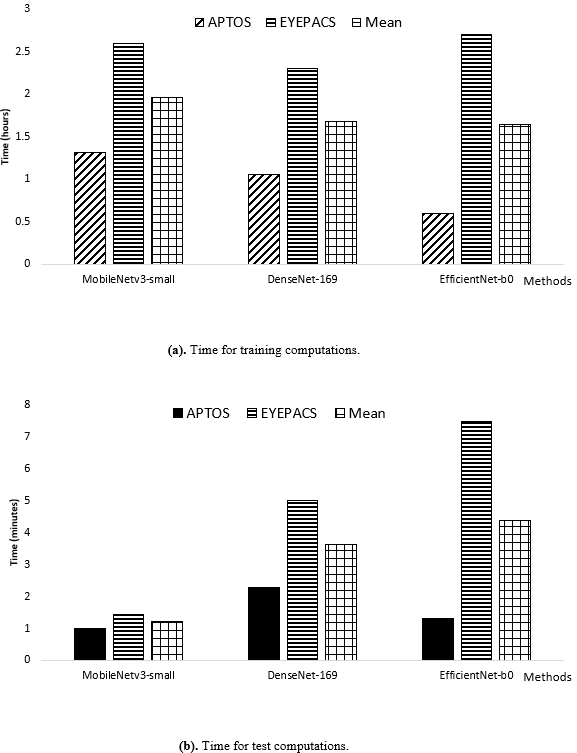}
  \caption{Computational complexity in terms of computational time for training and test phases.}
  \label{fig6}
  \end{center}
  
\end{figure*}

\subsection{Computational Complexity }
Fig. \ref{fig6}(a) illustrates the computational complexity of our approach during the training phase on both the APTOS and EYEPACS datasets.
The x-axis in Fig. \ref{fig6}(a) represents the backbone networks, which are MobileNetV3-small, Densenet-169, and EfficientNet-b0.
On the y-axis, the time consumed is depicted in hours. During the
training stage, the resolution of the images was set to 512×512 pixels.
On the APTOS dataset, the EfficientNet-b0 consumed 0.60 hours, followed
by the DenseNet-169 and MobileNetV3-small, which consumed 1.06 and 1.32
hours, respectively. Hence, on the APTOS dataset, the EfficientNet-b0
requires the least time, while the MobileNetV3-small requires the highest
time. On the EYEPACS dataset, which contains over 35,000 images, the
DenseNet-169 consumed 2.30 hours during the training phase, while the
MobileNetV3-small and EfficientNet-b0 consumed 2.60 and 2.30 hours,
respectively. As indicated in Fig. \ref{fig6}(a), on the EYEPACS dataset, the
DenseNet-169 consumed the least time during the training phase.
Moreover, on average, the EfficientNet-b0 requires 1.65 hours during
training phase on both the datasets, while the DenseNet-169 and
MobileNetV3-small, on average, consumed 1.68 and 1.96 hours,
respectively. Therefore, during the training phase, our study indicates
that the EfficientNet-b0 is computationally the most effective, followed
by the DenseNet-169 and the MobileNetV3-small.

Fig. \ref{fig6}(b) shows the test time of the dataset in minutes. For the 512×512
pixels image resolution, on the APTOS dataset, the MobileNetV3-small
consumed nearly 1 minute to test the whole dataset, while the
DenseNet-169 and EfficientNet-b0 consumed 2.30 and 1.30 minutes,
respectively. So, on the APTOS dataset, the MobileNetV3-small consumed
the least time. On the EYEPACS dataset, the EfficientNet-b0 consumed the
highest 7.50 minutes to test this dataset, followed by the
DenseNet-169 and MobileNetV3-small, which consumed 5 minutes and 1.45 minutes, respectively. Moreover, as indicated in Fig. \ref{fig6}(b), the MobileNetV3-small consumed 1.22 minutes during the test
phase on both datasets. Furthermore, the DenseNet-169 and
EfficientNet-b0, on average, consumed 3.65 and 4.40 minutes, respectively.
Therefore, during the test phase, our study indicates that the
MobileNetV3-small is computationally most efficient, followed by the
DenseNet-169 and the EfficientNet-b0, respectively.

\section{Conclusion}
\label{sec:conclusion}
In this paper, we presented a deep learning pipeline for DR grading. The proposed pipeline incorporates CABNet, a novel approach that merges the CAB and the GAB modules. We trained the CABNet holistically for DR grading, leveraging an attention module to learn distinctive features. The proposed approach classifies retinal images into five grades of DR. The approach employs aCNN-based model and an attention module. We have also explored MobileNetV3-small, DenseNet-169, andEfficientNet-b0 as the backbone model. We evaluated the proposedmethod on the APTOS and the EYEPACS datasets. Experimental resultsrevealed that on the APTOS dataset, the DenseNet-169 achieved a meanaccuracy of 83.20\%, followed by the MobileNetV3-small andEfficientNet-b0, which achieved 82\% and 80\%, respectively. On theEYEPACS dataset, the EfficientNet-b0 yielded a mean accuracy of 80\%. Onthis dataset, the DenseNet-169 and MobileNetV3-small yielded75.43\% and 76.68\% accuracies, respectively. In addition, the F1-scoreand Kappa-score are also reported. Our proposed method with the MobileNetV3-small consumed 0.90 million parameters on the EYEPACS dataset. Overall, the proposed method produces results on par with recently reported works. Utilizing synthetic datasets for initial deep model training, followed by fine-tuning authentic retinal fundus data, could enhance the overall performance in grading diabetic retinopathy. Hence, in the future, we aim to utilize neural diffusion models to generate high-quality retinal fundus images that can help in creating a generalized deep learning model for DR grading. This approach is expected to enhance model robustness and improve performance across diverse clinical scenarios.

\backmatter

\section*{Declarations}
\begin{itemize}
\item Funding: No funding was received for conducting this study.
\item Competing interests: No potential competing interest was reported by the author(s).
\item Ethics approval: Not applicable. 
\item Consent to participate: Not applicable.
\item Consent for publication: Not applicable.
\item Code availability: Not applicable.
\item Author contribution
\item Ethics approval: Not applicable. 
\item Consent to participate: Not applicable.
\item Consent for publication: Not applicable.
\item Availability of data and materials: All data analysed during this study is publicly available from https://www.kaggle.com/datasets/mariaherrerot/aptos2019 and https://www.kaggle.com/c/diabetic-retinopathy-detection.
\item Code availability: Not applicable.
\item Authors' contributions: Z. M. and H. A. designed the research. A. H. conducted the experiments. A. H., R. Q., H. A. the analysis and interpretation of the results. H. A. and Z. M. supervised the work. A. H. and Z. M. wrote the manuscript. H. A. and R. Q. revised the manuscript. All authors reviewed the manuscript. 
\end{itemize}

\bibliography{main}

\end{document}